\documentclass[twocolumn,10p]{amsart}

\usepackage[utf8]{inputenc}
\usepackage[english]{babel} 
\usepackage{graphicx,color,comment}  
\usepackage[dvipsnames]{xcolor} 
\usepackage{geometry,hyperref}
\usepackage{amsmath,amssymb,amsfonts,amsthm} 
\usepackage{mathtools,stmaryrd}  
\definecolor{light-gray}{gray}{0.95}
\usepackage[color=light-gray]{todonotes}
\usepackage{tikz-cd}  
\usepackage[all,cmtip]{xy} 
\usepackage{enumerate}

\newcommand{\dd}{\partial}
\newcommand{\gh}{\mathrm{gh}}
\newcommand{\g}{\mathfrak{g}}
\newcommand{\mr}{\mathrm}
\newcommand{\mc}{\mathcal}
\newcommand{\X}{\mathcal{X}}
\newcommand{\F}{\mathcal{F}}
\newcommand{\ZZ}{\mathbb{Z}}
\newcommand{\LL}{\mathcal{L}}
\newcommand{\NN}{\mathcal{N}}
\newcommand{\bt}{\bullet}
\newcommand{\ra}{\rightarrow}

\newcommand{\RR}{\mathbb{R}}

\newcommand{\ddd}{\mathrm{d}}

\DeclareMathOperator{\Map}{Map}

\theoremstyle{remark}
\newtheorem{remark}{Remark}[section]
\theoremstyle{plain}
\newtheorem{thm}[remark]{Theorem}
\theoremstyle{definition}
\newtheorem{definition}[remark]{Definition}
\newtheorem{example}[remark]{Example}

\title{BV Quantization - Encyclopedia of Math Phys}
\author{Alberto S. Cattaneo}
\address{Institut f\"ur Mathematik, Universit\"at Z\"urich, Winterthurerstrasse 190, 8057 Z\"urich, Switzerland}
\email{cattaneo@math.uzh.ch}
\author{Pavel Mnev}
\address{Department of Mathematics, University of Notre Dame, Notre Dame, Indiana 46556, USA}
\address{St. Petersburg Department of V. A. Steklov Institute of Mathematics of the Russian Academy of Sciences, Fontanka 27, St. Petersburg, 191023 Russia}
\email{pmnev@nd.edu}
\author{Michele Schiavina}
\address{Department of Mathematics, University of Pavia, Via Ferrata 5, 27100 Pavia, Italy}
\address{INFN Sezione di Pavia, via Bassi 6, 27100 Pavia, Italy}
\email{michele.schiavina@unipv.it}

\thanks{This article is commissioned by the Encyclopedia of Mathematical Physics, edited by M. Bojowald and R.J. Szabo, to be published by Elsevier \cite{Encyclopedia}. A.S.C. acknowledges partial support of the SNF
Grant No.\ 200020\_192080, of the Simons Collaboration on Global Categorical Symmetries, and of the COST Action 21109 (CaLISTA). This research was (partly) supported by the NCCR SwissMAP, funded by the Swiss National Science Foundation.}
\keywords{BV formalism. Gauge theories. Odd symplectic geometry. AKSZ construction.}

\begin{document}

\begin{abstract}
This note gives an overview of the BV formalism in its various incarnations and applications. 
\end{abstract}

\maketitle

Key points 
\begin{itemize}
\item Historical overview
\item BV quantization (central idea)
\item Zinn-Justin's effective action
\item Geometric interpretation
\item Relation to cyclic $L_\infty$ algebras and homotopy transfer
\item Various implementations in field theory
\end{itemize}

\section{Introduction}

In \cite{BV81,BV83} Batalin and Vilkovisky proposed a method of quantization of Lagrangian gauge systems generalizing the BRST \cite{BRS,T} approach to the case of open gauge symmetry algebra, i.e., to the case when the commutator of infinitesimal gauge transformations is a gauge transformation only on-shell (see, e.g., \cite{HenneauxTeitelboim}). 

Since the original papers, the BV formalism turned out to be very useful in the study of supersymmetric and topological field theories, see, e.g., \cite{AKSZ,CF,Getzler}. 

There are surprising cases of BV structure emerging in settings distant from quantum field theory, e.g., in Chas and Sullivan's string topology 
\cite{Chas-Sullivan}. 

The BV formalism is intimately tied with the geometry of odd-symplectic manifolds, the structure of integral forms and integration of half-densities over Lagrangian submanifolds, and it is ultimately an application of the theory of (exact) Gerstenhaber algebras \cite{YKSGersten}.

In this entry, we begin with a historical over\-view of the formalism to outline the motivation, followed by a description of its geometric and algebraic foundations, in finite dimensions. We conclude by providing a glimpse into several different ways one can extend the finite-dimensional insights to the actual scenario of field theory, showcasing a number of different current research directions.

\section{BV quantization: the idea}
In this section we sketch the original Batalin--Vilkovisky construction \cite{BV81}. 

\textbf{Input.} As input, consider a Lagrangian gauge system defined by the classical action functional $S_{cl}(\phi)$---a function
of classical fields $\phi^i$ (local coordinates on the manifold of classical fields $X$, assumed here to be finite-dimen\-sional), with infinitesimal gauge transformations given by vector fields $v_a=v_a^i(\phi)\frac{\dd}{\dd \phi^i}$ on $X$ preserving $S_{cl}$. Note that the vector fields $v_a$ are automatically tangent to the critical locus of $S_{cl}$ (i.e., the 
locus
of fields satisfying the Euler--Lagrange equations).

\textbf{Step I.} Extend $S_{cl}(\phi)$ to a function $S(\Phi,\Phi^+)$ (the \emph{master action} or \emph{BV action})
satisfying the \emph{classical master equation} (CME)
\begin{equation}\label{CME}
(S,S)=0.
\end{equation}
Here:
\begin{itemize}
    \item $\Phi^I$ stands for classical fields $\phi^i$ and anticommuting \emph{ghost fields} $c^a$ associated with gauge symmetries. For each field $\Phi^I$ one has an associated \emph{antifield} $\Phi_I^+$ of opposite parity (i.e., an anti-classical field $\phi_i^+$ or an antighost $c^+_a$). One assigns a $\mathbb{Z}$-grading  (called \emph{ghost number}) as follows:
    $\gh(\phi^i)=0$, $\gh(c^a)=1$, $\gh(\Phi^+_I)=-1-\gh(\Phi^I)$.
    \item In (\ref{CME}), the symbol $(\ ,\ )$ is the \emph{antibracket}\footnote{Or ``BV bracket,'' or ``odd Poisson bracket,'' or ``Gerstenhaber bracket.''} defined by
    \begin{multline*}
        (f,g)=\sum_I (-1)^{\gh(\Phi^I)}\\ f \left(\frac{\overleftarrow\dd}{\dd \Phi^I} \frac{\overrightarrow\dd}{\dd \Phi^+_I}-\frac{\overleftarrow\dd}{\dd \Phi^+_I} \frac{\overrightarrow\dd}{\dd \Phi^I}\right) g.
    \end{multline*}
    \item The master action $S$ is even with ghost number equal to $0$ and is related to 
    the classical gauge system ($S_{cl},\{v^a\})$ as follows:
    \begin{multline}\label{BV ansatz}
        S(\Phi,\Phi^+)=S_{cl}(\phi)\\+\phi^+_i v^i_a(\phi) c^a +\frac12 f_{ab}^c(\phi) c^+_c c^a c^b \\
        +\sum_{k\geq 2} \Phi^+_{I_1}\cdots \Phi^+_{I_k} S^{I_1\cdots I_k}(\Phi)
    \end{multline}
    for some structure functions $f^a_{bc}(\phi)$ and some functions $S^{I_1\cdots I_k}(\Phi)$.
\end{itemize}

For further details on this construction, see \cite{HenneauxEnc}.

\begin{remark}
    There is a different approach where the input is just the function $S_{cl}$ on $X$ (and symmetries $v_a$ are not part of the input). Then the aim is to construct an extension of $S_{cl}$ to a solution $S$ of CME on the extended space $\X$ satisfying a ``properness'' (or ``maximal nondegeneracy'') axiom: for $(\Phi,\Phi^+)$ a critical point of $S$, the kernel of the 
    Hessian $\mr{ker}\,\dd^2 S(\Phi,\Phi^+)$ is a Lagrangian subspace of the tangent space $T_{\Phi,\Phi^+}\X$.\footnote{As a consequence of CME, the kernel of the Hessian is automatically coisotropic. ``Lagrangian'' and ``coisotropic'' here refers to the odd symplectic structure $\omega=\delta \Phi^I \wedge \delta \Phi^+_I$.} 
    The extended space $\X$ and the function $S$ are constructed from the Koszul-Tate resolution of the algebra of functions on the critical locus of $S_{cl}$. See \cite{FK} for details and existence-uniqueness theorem in this setup. Note that in this approach symmetries $v_a$ are recovered from the Koszul-Tate differential.
\end{remark}

\begin{remark}
If the gauge transformations $v^a$ are closed under Lie bracket, one does not need quadratic and higher terms in $\Phi^+$ in (\ref{BV ansatz})---this is the case of gauge theories that can be treated using the BRST formalism. However, for an \emph{open} gauge symmetry algebra, i.e., in the situation when $[v_a,v_b]=f^c_{ab}(\phi) v_c+\cdots$ with $\cdots$ a term vanishing only on-shell (i.e., on the critical locus of $S_{cl}$), one needs $k\geq 2$ terms in (\ref{BV ansatz}). These are the cases when the BV construction is necessary for quantization, while BRST is insufficient.
\end{remark}

\begin{remark} The classical master equation implies that the odd derivation \[Q=(S,-)\] squares to zero;\footnote{Other traditional notations for it are $s$ and $\delta$.} $Q$ is known as the (classical) BRST operator.
\end{remark}

\textbf{Step II (quantization).} Extend $S$ by adding corrections in powers of $\hbar$ to get a solution $S_\hbar=S+\hbar S^{(1)}+\hbar^2 S^{(2)}+\cdots$ of the \emph{quantum master equation} (QME)
\begin{equation*}
    \Delta e^{\frac{i}{\hbar}S_\hbar} =0,
\end{equation*}
or, equivalently,
\begin{equation*}
    \frac12(S_\hbar,S_\hbar)-i\hbar \Delta S_\hbar =0,
\end{equation*}
where the operator
\begin{equation*}
    \Delta=\frac{\dd^2}{\dd \Phi^I \dd\Phi^+_I}\end{equation*}
is the so-called BV Laplacian. The QME is equivalent to a sequence of equations
\begin{multline*}
(S,S)=0,\; (S,S^{(1)})-i\Delta S=0,\\(S,S^{(2)})+\frac12(S^{(1)},S^{(1)})-i\Delta S^{(1)}=0,\cdots
\end{multline*}
\begin{remark}\label{rem-anomaly}
    Given a solution $S$ of the CME, one can construct corrections $S^{(\geq 1)}$ one-by-one, solving the chain of equations above, provided that certain obstructions in the first cohomology of $Q$ vanish (if this extension is not possible, the theory has gauge anomalies; see \cite{ChiaffrinoSachs} for a discussion).
\end{remark}

The partition function of the quantum system is then defined as
\begin{equation}\label{BV integral}
    Z=\int e^{\frac{i}{\hbar}S_\hbar(\Phi^I,\Phi_I^+=\frac{\dd \Psi}{\dd \Phi^I})} \prod_I D\Phi^I
\end{equation}
where the odd ($\gh=-1$) function $\Psi(\Phi)$ is the so-called \emph{gauge-fixing fermion}. The main observation underlying the Batalin--Vilkovisky formalism is that the integral (\ref{BV integral}) is invariant under deformations of $\Psi$ (as a consequence of the QME for $S_\hbar$).

The integral (\ref{BV integral}) is called a BV integral and defines the gauge-fixed functional integral for the original gauge system.  For $\Psi$ satisfying a certain nondegeneracy condition,\footnote{The exponent in (\ref{BV integral}) should have nondegenerate critical points.} one can express the stationary phase asymptotics of this integral as a sum over Feynman diagrams.

\begin{remark}
The fields $\Phi^I=\psi^i,c^a$ so far were nonnegatively graded, so it is unclear how to construct a gauge-fixing fermion $\Psi(\Phi)$ with $\gh=-1$. The idea is to extend the fields $\Psi$ by \emph{auxiliary fields} $b_a$ and $\lambda_a$ (with $\gh=-1,0$, respectively)\footnote{The field $b_a$ is also denoted $\bar{c}_a$ and is known as the ``second Faddeev--Popov ghost'' or as the ``antighost.'' When one chooses the latter terminology, one then calls $c^{+}_a$ the ``ghost antifield.''} satisfying $Qb_a=\lambda_a$. One then has the corresponding antifields $b^{+a}$ and $\lambda^{+a}$, and one adds the term $\lambda_a b^{+a}$ to $S$. With this amendment (sometimes called \emph{nonminimal} BV formalism), one can consider gauge-fixing fermions (for instance) of the form $\Psi=b_a F^a(\phi)$. With this particular choice, the gauge-fixing for the physical fields is given by the set of equations $F^a(\phi)=0$.
\end{remark}

\textbf{Observables.} Assuming $Z\not=0$, one is also interested in computing expectation values of functions $\mathcal{O}$ via
\[
\langle O\rangle \coloneqq \frac1Z
\int e^{\frac{i}{\hbar}S_\hbar} 
\ \mathcal{O}
\prod_I D\Phi^I
\]
with the assignment $\Phi_I^+=\frac{\dd \Psi}{\dd \Phi^I}$. To apply the BV theorem
and make sure that $\langle O\rangle$ is invariant under deformations of $\Psi$, we have to assume $\Delta (e^{\frac{i}{\hbar}S_\hbar}\mathcal{O}) =0$ in addition to the QME. This is equivalent to
\[
(S_\hbar,\mathcal{O}) - i\hbar
\Delta \mathcal{O}   = 0.
\]
A function $\mathcal{O}$ satisfying this equation is called a BV observable. If $\mathcal{O}$ satisfies $(S,\mathcal{O})=0$, it is called a classical BV observable. A classical BV observable depending only on the fields, but not on the antifields, formally satisfies $\Delta\mathcal{O}=0$, so it is a BV observable as well, assuming that $S_\hbar=S$.\footnote{Or, more generally, assuming that $\hbar^{\geq 1}$ terms in $S_\hbar$ do not depend on the antifields.}

\begin{example}
For Yang--Mills theory on a (pseudo--)Riemannian $n$-manifold $M$ with structure Lie group $G$ with Lie algebra $\g$ equipped with invariant nondegenerate inner product $\langle,\rangle$, the master action is
\begin{multline*}
S=S_\hbar= \int_M \Bigg(  \frac12 \langle F_A \stackrel{\wedge}{,} * F_A \rangle\\
+\langle A^+  ,\ddd_A c\rangle +\frac12 \langle c^+ ,[c,c]\rangle + \langle \lambda, b^+\rangle \Bigg).
\end{multline*}
Here $A\in \Omega^1(M,\g)$ is the gauge field (connection in a trivial $G$-bundle over $M$, with curvature $F_A=dA+\frac12[A,A]$), $c\in \Omega^0(M,\g)_{1}$ is the ghost,\footnote{The subscript  stands for the ghost number.} $A^+\in \Omega^{n-1}(M,\g)_{-1}$ is the antifield, $c^+\in \Omega^{n}(M,\g)_{-2}$ is the antighost; $b\in \Omega^{n}(M,\g)_{-1}, \lambda\in \Omega^n(M,\g)_0$ are the auxiliary fields and \\
$b^+\in \Omega^0(M,\g)_{0},  \lambda^+\in \Omega^0(M,\g)_{-1}$ are their corresponding antifields. Then imposing, e.g., the Lorenz gauge $\ddd^*A=0$ corresponds to choosing $\Psi=\int_M \langle b, \ddd^*A\rangle$. The exponent in (\ref{BV integral}) becomes
$$ \int_M \left(\frac12 \langle F_A \stackrel{\wedge}{,} *F_A \rangle +\langle \lambda, \ddd^*A\rangle + \langle b, \ddd^* \ddd_A c\rangle \right) $$
and coincides with the Faddeev--Popov gauge-fixed action for Yang--Mills theory. The integral (\ref{BV integral}) then gives rise to the standard Feynman diagrams for Yang--Mills theory \cite{FP}. A gauge-invariant functional $\mathcal{O}$ of the gauge field $A$ is then an example of BV observable.
\end{example}

\begin{remark}\label{rem: higher ghosts}
    The construction outlined in this section also applies to the case when there are linear dependencies between gauge symmetries $v^a$ (i.e., to the case of reducible gauge symmetry), \cite{BV83}. In this case one needs to adjoin \emph{higher ghosts} (with $\gh=2$) to the fields $\Phi$ (and the corresponding antifields to $\Phi^+$). Likewise, one might have a tower of reducibility (corresponding to a 
    homological resolution of gauge symmetry by ``free'' objects\footnote{Spaces of sections of vector bundles over the spacetime manifold, in the context of local field theory.}), necessitating the introduction of a tower of higher ghosts of increasing ghost number.
    
    \textbf{Example:}
    $p$-form electrodynamics, 
    \[
    S_{cl}=\int_M \frac12 \ddd A^{(p)}\wedge *\ddd A^{(p)},
    \] 
    with $\phi=A^{(p)}$ a $p$-form. One has gauge symmetry $A^{(p)}\mapsto A^{(p)}+\ddd A^{(p-1)}$, with generators $A^{(p-1)}$ and $A^{(p-1)}+\ddd A^{(p-2)}$ acting by the same transformation. Ultimately, one has a tower 
    \[
    \Omega^0(M)\xrightarrow{\ddd} \Omega^1(M)\xrightarrow{\ddd}\cdots \xrightarrow{\ddd} \Omega^{p-1}(M)\xrightarrow{\ddd} \Omega^{p}(M)
    \]
    and the corresponding tower of higher ghosts.
\end{remark}

\subsection{Zinn-Justin's effective action}
One interesting application of the BV formalism, which actually predates its full definition, is the construction by Zinn-Justin \cite{{ZinnJustin}} of an effective action that satisfies the classical master equation (CME). This is the starting point of one of the renormalization techniques in field theory. We follow the presentation in \cite{Anselmi}, where this construction is presented in full generality.

The central idea consists in introducing sources $J_I$ for the fields $\Phi^I$ and sources $K^+_I$ for the BRST variations $Q\Phi^I$. Namely, one defines
\[
Z(J,K^+)=\int e^{\frac{i}{\hbar}(S_\hbar+J_I\Phi^I+K^+_IQ\Phi^I)} \prod_I D\Phi^I
\]
with $\Phi_I^+=\frac{\dd \Psi}{\dd \Phi^I}$.\footnote{Equivalently, one can view the $K^+_I$s as parametrizing linear deformations of $\Psi$. Namely, one has
\[
Z(J,K^+)=\int e^{\frac{i}{\hbar}(S_\hbar+J_I\Phi^I)} \prod_I D\Phi^I
\]
with $\Phi_I^+=\frac{\dd \Tilde\Psi}{\dd \Phi^I}$ and $\Tilde\Psi=\Psi+K^+_I\Phi^I$.
}
One writes $Z=e^{\frac i\hbar W}$ and defines the effective action $\Gamma(K,K^+)$ as the Legendre transform of $W(J,K^+)$ with respect to $J$. It then follows that, if $S$ satisfies the QME, then $\Gamma$ satisfies the CME with respect to the fields $K$ and antifields $K^+$.

If the integral is computed perturbatively by the saddle-point approximation, it then happens that $\Gamma$ is the generating function of 1PI Feynman diagrams. Moreover, $\Gamma(K,K^+)= S(K,K^+)+O(\hbar)$. This is quite useful in field theory. One chooses a regularization to perform the integrals and modifies the action $S$ by adding counterterms. The so obtained action is now of the form $\Sigma=\sum_{n=0}^\infty \hbar^n \Sigma^{(n)}$, with $\Sigma^{(0)}=S$.
One assumes by induction that, up to order $\hbar^n$, one has been able to find the counterterms that, up to order $\hbar^n$, $i)$ make the effective action $\Gamma$ finite and $ii)$ make $\Sigma$ satisfy the QME. It then turns out that divergent terms in $\Gamma$ at order $\hbar^{n+1}$ are local and $Q$-closed.
One can then reabsorb them by changing $\Sigma^{(n+1)}$ via $i)$ a redefinition of the coupling constants and $ii)$ a BV canonical transformation, completing the induction.\footnote{To be precise, there is an interplay between the classical and the quantum master equation. Up to a certain order in $\hbar,$, the effective action satisfies the CME if the action satisfies the QME. The inductive step produces a solution of the CME, but this has to be extended to a solution of the QME. For this to happen, one assumes there are no anomalies (see Remark~\ref{rem-anomaly}).} One also sees that the possible counterterms are classified by the local $Q$-cohomology\footnote{If this turns out to be finite-dimen\-sional, then one needs only finitely many coupling constants to renormalize.} \cite{FHST}.

\section{Geometric interpretation of the BV formalism}
In this section we outline the geometric viewpoint on the BV formalism due to A. Schwarz \cite{Schwarz}.

A supermanifold\footnote{
``Super'' means that local coordinates are $\ZZ_2$-graded and are commuting or anticommuting according to the $\ZZ_2$-degree.
} $\X$ is said to be a \emph{$P$-manifold} if it is equipped with an odd symplectic structure $\omega\in \Omega^2(\X)_\mr{odd}$---a nondegenerate closed odd 2-form. If additionally $\X$ is equipped with a compatible volume element $\mu$, then $\X$ is called an $SP$-manifold. Compatibility means that there is an atlas of local Darboux charts\footnote{We denote by $(x,\xi)$ a generic Darboux chart. This is to differentiate w.r.t.\ the notation $(\Phi^I_,\Phi^+_I)$, which instead typically refers to a particular type of chart where the ``fields'' $\Phi$ have nonnegative ghost number.} $(x^i,\xi_i)$ such that locally $\omega=\ddd x^i\wedge \ddd \xi_i$ ($\xi_i$ has the opposite parity of $x^i$) and $\mu=\prod_i Dx^i\,D\xi_i$---the coordinate Berezinian.

For a function $f\in C^\infty(\X)$ on a $P$-manifold, one has the Hamiltonian vector $X_f$ on $\X$ of opposite parity, defined by
$\iota_{X_f}\omega=\ddd f$. One then defines the odd Poisson bracket on $C^\infty(\X)$ by
$$(f,g)\coloneqq (-1)^{\epsilon(f)} X_f(g)$$
where $\epsilon(f)\in \ZZ_2$ is the parity of $f$.
Using the $S$-strucure, one defines the BV Laplacian by 
\begin{equation}\label{BV-Schwarz Laplacian}
\Delta_\mu(f)\coloneqq\frac12 \mr{div}_\mu X_f
\end{equation}
where $\mr{div}_\mu$ is the divergence of a vector field with respect to $\mu$.\footnote{One defines $\mr{div}_\mu v$ by $\int_\X  \mu\, v(\rho) = - \int_\X \mu\, \mr{div}_\mu(v) \rho $ for any test function $\rho$.} Compatibility of $\mu$ with $\omega$ implies in particular the property $\Delta_\mu^2=0$. The definition (\ref{BV-Schwarz Laplacian}) is due to Khudaverdian \cite{Khudaverdian91}.

\begin{remark}\label{rem: odd cotangent bundle}
    For any supermanifold $\mc{N}$, the cotangent bundle with reversed parity of the fiber $\X=\Pi T^* \mc{N}$ is canonically a $P$-manifold. If $\mc{N}$ carries a volume element $\nu$, then $\X$ is also an $SP$-manifold, with $\mu=\nu^2$. Locally, if $x^i$ are coordinates on $\mc{N}$ and $\nu=\rho(x)\prod_i Dx^i$ with $\rho$ some density function, then $\omega=\ddd x^i\wedge \ddd \xi_i$ and $\mu=\rho(x)^2 \prod_i Dx^i D\xi_i$. 
    
    A theorem of Schwarz \cite{Schwarz}, building on older results of Batchelor \cite{Bat79}, says that any $P$-manifold can be globally written\footnote{That is, there is some symplectomorphism between $\X$ and $\Pi T^*N$.} as 
    \begin{equation}\label{Pi T^*N}
    \X=\Pi T^*N
    \end{equation} 
    for an ordinary (purely even) manifold $N$.\footnote{One can think of this statement as a global Darboux theorem in odd symplectic geometry. For an even symplectic form $\omega$, a similar statement does not hold.}
\end{remark}

\textbf{Lagrangian submanifolds.} 
A submanifold $i\colon\LL\hookrightarrow \X$ of a $P$-manifold $\X$ of even$|$odd dimension $(n|n)$ is said to be \emph{Lagrangian} if it is isotropic $i^*\omega=0$ and maximal, i.e., has dimension $(k|n-k)$ for some $k$.

The two main examples of Lagrangian submanifolds in $\X=\Pi T^*\mc{N}$ are the following.
\begin{enumerate}[(i)]
    \item For $C\subset \mc{N}$ a submanifold, the conormal bundle $\Pi N^*C$ (its fiber over $x\in\mc{N}$ is the parity-reversed annihilator of the tangent space $T_x C$ in $T^*_x\mc{N}$) is a Lagrangian submanifold of $\X$.
    \item For a function $\Psi\in C^\infty(\mc{N})_\mr{odd}$ (``gauge-fixing fermion''), one has that
$$
\mr{graph}(d\Psi) = \Big\{(x,\xi)\;|\; \xi_i = \frac{\dd \Psi(x)}{\dd x^i}\Big\}
$$
is a Lagrangian submanifold of $\X$. One can think of this example as a deformation of the zero-section Lagrangian $\NN\subset \Pi T^*\NN$, where $\Psi$ is the parameter of the deformation.
\end{enumerate}

A theorem of Schwarz asserts that any Lagrangian submanifold $\LL$ of a $P$-manifold (\ref{Pi T^*N}) can be obtained by the conormal bundle construction (i) above $\LL'=\Pi N^*C$, followed by a deformation (ii) by a gauge-fixing fermion $\Psi\in C^\infty(\LL')_\mr{odd}$.\footnote{Here another result of Schwarz is implicitly used: a tubular neighborhood of a Lagrangian $\LL'$ in $\X$ is symplectomorphic to 
a tubular neighborhood of the zero-section in
$\Pi T^*\LL'$.}

\textbf{BV integrals.}
Given a $P$-manifold $\X$ and a Lagrangian submanifold $\LL$, one can canonically construct a volume element $\sqrt{\mu}|_\LL$ on $\LL$ out of any volume element $\mu$ on $\X$. If $\LL$ is given in local Darboux coordinates $(x^i,\xi_i)$ by $\xi=0$, and $\mu=\rho(x,\xi)\prod_i Dx^i D\xi_i$, then $\sqrt{\mu}|_\LL=\sqrt{\rho(x,0)}\prod_i Dx^i$.

Now let $(\X,\omega,\mu)$ be an $SP$-manifold. A BV integral in this setting is an integral of the form
\begin{equation}\label{BV integral (Schwarz)}
\int_\LL f \sqrt{\mu}|_\LL
\end{equation} 
where $\LL\subset \X$ is a Lagrangian submanifold and $f\in C^\infty(\X)$ is a function.
\begin{thm}[Schwarz]\label{thm: Schwarz BV-Stokes}
    \begin{enumerate}[(a)]
        \item If $\Delta_\mu f=0$, then the BV integral (\ref{BV integral (Schwarz)}) is unchanged under continuous deformations of $\LL$ in the class of Lagrangian submanifolds of $\X$.
        \item If $f=\Delta_\mu g$ for some function $g$ on $\X$, then the BV integral (\ref{BV integral (Schwarz)}) vanishes.
    \end{enumerate}
\end{thm}

\begin{remark}
    So far in this section we were describing the structure in terms of $\ZZ_2$-grading. In many examples it can be refined to a $\ZZ$-grading (by ghost number), so that $\epsilon(f)=\gh(f) \bmod 2$. Then $\omega$ conventionally has degree $\gh=-1$, while $(,)$ and $\Delta_\mu$ have degree $\gh=+1$.
\end{remark}

\subsection{BV algebras}
A BV algebra is a supercommutative unital algebra $(A,\cdot,1)$ equipped with an odd second-order derivation $\Delta$ satisfying $\Delta^2=0$ and $\Delta(1)=0$. As a consequence, $A$ carries an odd Poisson bracket $(\ ,\ )$ defined via
$$\Delta(f\cdot g)=\Delta(f)\cdot g+(-1)^{\epsilon(f)} f\cdot \Delta(g)+(-1)^{\epsilon(f)} (f,g) $$
where $f,g\in A$.

In particular $(A,\cdot, (,))$ is a Gerstenhaber (or ``odd Poisson'') algebra.

\textbf{Examples.}
\begin{enumerate}
    \item For $(\X,\omega,\mu)$ an $SP$-manifold, the algebra of functions $A=C^\infty(\X)$ is a BV algebra, via the construction (\ref{BV-Schwarz Laplacian}).
    \item For $N$ a manifold equipped with a volume form $\nu$, the algebra of multivector  fields $A=\mc{V}^\bt(N)=\Gamma(N,\wedge^\bt TN)$ is a BV algebra, with $\Delta=\mr{div}_\nu$ the divergence operator sending a $k$-vector to a $(k-1)$-vector. The corresponding odd Poisson bracket $(\ ,\ )$ is the Schouten--Nijenhuis bracket $[\ ,\ ]_\text{SN}$ of multivector fields (an extension of the Lie bracket of vector fields to a biderivation of the algebra of multivectors).\footnote{In fact, this example is a special case (or rather a reinterpretation) of the previous one, where $\X=\Pi T^*N$, with $\mu$ as in Remark \ref{rem: odd cotangent bundle}.} In this example $N$ can be replaced with a supermanifold.
\end{enumerate}

\begin{remark} The following observation is due to Witten \cite{Witten}. For $N$ an $n$-manifold equipped with a volume form $\nu$, one has an isomorphism between the de~Rham complex of $N$ and multivectors equipped with the BV Laplacian $\Delta=\mr{div}_\nu$:
\begin{equation*}
    \underbrace{C^\infty(\Pi TN)  }_{\Omega^\bt(N)} 
    \xrightarrow{\sigma} \underbrace{C^\infty(\Pi T^* N)}_{\mc{V}^\bt(N)} 
\end{equation*}
The map here is the fiberwise \emph{odd Fourier transform}
\begin{multline*}
f(x,\psi)\mapsto  \check{f}(x,\xi)
=\int_{\Pi T_x N}D\psi\;e^{\langle \psi,\xi\rangle}f(x,\psi) 
\end{multline*} 
where $\psi^i=\ddd x^i$ are fiber coordinates on $\Pi T_x N$, and the fiber Berezinian $D \psi$ is determined by the volume form $\nu$ at $x\in N$. The inverse map $\sigma^{-1}$ can be described as contraction of multivectors with $\nu$, mapping a $k$-multivector to an $(n-k)$-form. The observation is that the odd Fourier transform $\sigma$ intertwines the de~Rham operator $\ddd=\psi^i\frac{\dd}{\dd x^i}$ and the divergence/BV Laplace operator. If $\nu$ is locally given by a constant density, then the latter is $\frac{\dd^2}{\dd x^i \dd\xi_i}$.
\end{remark}

\subsection{Canonical BV Laplacian on half-den\-sities}
The BV Laplacian (\ref{BV-Schwarz Laplacian}) on functions on a $P$ manifold depends on a choice of compatible volume element $\mu$. For the moment, we denote it $\Delta_\mu$. 

H. Khudaverdian \cite{Khudaverdian} observed that on the space $\mr{Dens}^{\frac12}(\X)$ of half-densities on a $P$-manifold $\X$ (sections of the line bundle of half-densities defined by transition functions $\left|\frac{\dd (x',\xi')}{\dd(x,\xi)} \right|^{\frac12}$), there exists a \emph{canonical} BV Laplacian $\Delta^\mr{can}$, independent of the choice of $\mu$. Locally in a Darboux chart, it is given by
\begin{multline*}
\Delta^\mr{can}\colon f\cdot  \prod_i(Dx^i)^{\frac12}(D\xi_i)^{\frac12} \mapsto \\
\mapsto \left(\frac{\dd^2}{\dd x^i \dd \xi_i} f\right)\cdot \prod_i(Dx^i)^{\frac12}(D\xi_i)^{\frac12}. 
\end{multline*}
Khudaverdian showed that this is a globally well-defined operator, and \v{S}evera \cite{Severa} gave a different, manifestly global, construction for it.\footnote{
\v{S}evera considered differential forms on a $P$-manifold as a bicomplex $\Omega^\bt(\X),d,\omega\wedge$, with de Rham differential and the second differential given by wedging with the odd symplectic form. The cohomology of the second differential turns out to be canonically isomorphic to half-densities on $\X$. Considering the spectral sequence of the bicomplex, \v{S}evera showed that the third sheet $E_3$ is given by $H_{\omega\wedge}(\Omega^\bt(\X))\cong \mr{Dens}^{\frac12}(\X)$ with the induced differential $d_3=d(\omega\wedge)^{-1}d$ being the canonical BV Laplacian.
} 

If one chooses a compatible volume element $\mu$ on a $P$-manifold $\X$, then one has an isomorphism
$$C^\infty(X)\xrightarrow{\cdot \sqrt\mu} \mr{Dens}^{\frac12}(\X)$$
which intertwines the operators $\Delta_\mu$ on functions and $\Delta^\mr{can}$ on half-densities: $\Delta_\mu f = \frac{1}{\sqrt{\mu}} \Delta^{\mr{can}} (\sqrt{\mu}f)$.

A half-density $\alpha$ on $\X$ restricted to a Lagrangian submanifold $\LL\subset \X$ yields a 1-density (or volume element) on $\LL$. Thus, one can consider BV integrals
\begin{equation} \label{BV integral 1/2-density}
    \int_\LL \alpha.
\end{equation}

Phrased in terms of half-densities, Theorem \ref{thm: Schwarz BV-Stokes} says: if $\X$ is a $P$-manifold, then
\begin{enumerate}[(a)]
    \item for $\alpha$ a $\Delta^\mr{can}$-closed half-density on $\X$, the BV integral (\ref{BV integral 1/2-density}) is invariant under Lagrangian deformations of $\LL\subset X$;
    \item for $\alpha$ a $\Delta^\mr{can}$-exact half-density, the BV integral (\ref{BV integral 1/2-density}) is zero.
\end{enumerate}

\begin{definition} 
    Let $(\X,\omega,\mu)$ be an $SP$-manifold with canonical BV laplacian $\Delta^\mr{can}$. A solution of the Quantum Master Equation (QME) is an element $S\in C^\infty(\X)_{\mr{even}}\llbracket\hbar\rrbracket$ such that
    \[
    \Delta_\mu e^{\frac{i}{\hbar} S} = 0 \iff (S,S) - i\hbar \Delta_\mu S=0.
    \]
    Two solutions $S_0,S_1$ of the QME are \emph{equivalent} if there exists a family $S_t\in C^\infty(\X)_{\mr{even}}\llbracket\hbar\rrbracket$ such that
    \[
    \frac{d}{dt} S_t= (S_t, R_t) - i\hbar \Delta_\mu R_t
    \]
    for $R_t\in C^\infty(\X)_{\mr{odd}}\llbracket\hbar\rrbracket$ and $S_{t=0}=S_0$ and $S_{t=1}=S_1$. $S_t$ is called a canonical $BV$ transformation with generator $R_t$. We denote the set of equivalence classes of solutions of the QME under canonical BV transformations by $\mr{sQME}/\!\!\sim$.
\end{definition}

This notion of canonical transformation is the infinitesimal version of a transformation $S\to S'$ with
\begin{equation}\label{can transf quantum}
e^{\frac{i}{\hbar}S'}=e^{\frac{i}{\hbar}S}+\Delta_\mu\left(e^{\frac{i}{\hbar}S}R\right)
\end{equation}

In this setting we have:
\begin{thm}[Batalin-Vilkovisky]
    Let $(\X,\omega,\mu)$ be an $SP$-manifold and $S\in \mr{sQME}$. The BV integral 
    \[
    Z= \int_{\LL} e^{\frac{i}{\hbar}S} \sqrt{\mu}
    \]
    is invariant under canonical transformations.
\end{thm}

\subsection{BV fiber integrals}\label{sec:BVfiberintegral}
An important variation on the theme of BV integrals applies to an $SP$-manifold $(\X,\omega,\mu)$ with product structure $\X=\X_1\times \X_2$, so that both factors are $SP$-manifolds $(\X_i, \omega_i, \mu_i)$, and $\omega = \omega_1\otimes 1 + 1\otimes \omega_2\equiv \omega_1\oplus \omega_2$ as well as $\mu = \mu_1\mu_2$. We can perform BV integration on, say, the $\X_2$ factor by choosing a Lagrangian submanifold $\LL\subset \X_2$. From the decomposition 
\[
\mr{Dens}^{\frac12}(\X) \simeq \mr{Dens}^{\frac12}(\X_1)\hat{\otimes} \mr{Dens}^{\frac12}(\X_2)
\]
we obtain the BV fiber integral:
\[
\mathcal{P}^{(\LL)}_* \colon \mr{Dens}^{\frac12}(\X) \to \mr{Dens}^{\frac12}(\X_1), 
\]
where $\mathcal{P}^{(\LL)}_* = \mathrm{id}\otimes \int_\LL$. (Note that this procedure generalizes to nontrivial odd symplectic fibrations $\mathcal{P}\colon\X\to \X_1$.) 

\begin{definition}
Given a solution of the QME $S\in \mathrm{sQME}$, an effective action for $S$ induced on $\X_1$ is an element $S_1\in C^\infty(\X_1)\llbracket\hbar\rrbracket$ such that 
\begin{equation}\label{fiber BV integral}
e^{\frac{i}{\hbar} S_1} \mu_1  = \mathcal{P}^{(\LL)}_* (e^{\frac{i}{\hbar}S} \mu).
\end{equation}
\end{definition}

\begin{thm}[BV fiber integral]\label{thm: fib BV int}
The BV fiber integral descends to a map
\[
    \mathcal{P}_*^{[\LL]}\colon\mr{sQME}(\X)/\!\!\sim\ \longrightarrow\ \mr{sQME}(\X_1)/\!\!\sim
\]
where $[\LL]$ is the class of $\LL$ modulo \emph{Lagrangian} homotopy. 
\end{thm}

BV fiber integrals have several applications. 
\begin{enumerate}
    \item They can be used to generate some nontrivial BV observables \cite{CattaneoRossi,Mnev15,Moshayedi}. For this we assume that, in addition to the BV action $S_1$ satisfying the QME, we have a function $S_2$, depending on some extra fields, such that $S_1+S_2$ satisfies the QME for all fields; equivalently, 
    \[
    \frac12(S_2,S_2) + (S_1,S_2) -i\hbar\Delta_\mu S_2 = 0,
    \]
    where $\Delta_\mu$ is the BV Laplacian including the extra fields. The choice of a gauge-fixing Lagrangian $\mathcal{L}$ in the BV space of extra fields  $\X_2$ allows for the construction of a BV observable $\mathcal{O}$ for the original theory via 
    \begin{equation*}
    \mathcal{O}=\int_{\mathcal{L}} e^{\frac i\hbar S_2}\mu.
    \end{equation*}
    \item They are at the basis of the compatibility of renormalization \`a la Wilson and the BV formalism \cite{Losev06,Mnev06,Costello07}. Namely, one assumes that the BV action, on the the space $\X=\X_1\times \X_2$, satisfies the QME. One then interprets elements of $\X_2$ as ultraviolet (UV) fields. The BV fiber integral produces an effective action on the space $\X_1$ of infrared (IR) fields which also satisfies the QME. More on this in Section \ref{sec:FieldTheory}. 
    This procedure can be iterated. 
    In some cases, one can proceed all the way to a space $\X_1$ describing the BV (cohomological) resolution of the moduli space of critical points of the action.
    \item They can also be used to produce equivalences between field theories, where the elements of $\X_2$ are interpreted as auxiliary fields. (See, e.g., \cite{BBH95, Henneaux,RocekZeitlin,CanepaCattaneoSchiavinaAKSZ,SimaoCattaneoSchiavina}.)
\end{enumerate}

\subsection{Solutions of the master equation and cyclic $L_\infty$ algebras.}

Consider a special case of a $P$-manifold $(\X,\omega)$, with $\X$ a $\ZZ$-graded \emph{vector space} (with $\ZZ_2$-grading induced via $\bmod\;2$ reduction) and with $\omega$ a \emph{constant} nondegenerate 2-form of degree $-1$. 
Let 
\begin{equation}\label{S=S2+S3+..}
S=S_2+S_3+\cdots
\end{equation}
be a polynomial  function on $\X$  (or, more generally, a power series in coordinates on $\X$) starting with the quadratic term; $S_n$ stands for the homogeneous component of $S$ of polynomial degree $n$. Assume that $S$ satisfies the classical master equation (\ref{CME}). Then $S$ gives rise to a cyclic $L_\infty$ algebra structure on the degree-shifted graded vector space $V\coloneqq\X[-1]$. Indeed, $\omega$ induces a degree $-3$ inner product $\langle,\rangle\colon V\otimes V \ra \RR$. For each $n\geq 1$ the Hamiltonian vector field generated by $S_{n+1}$ induces (via degree shift and dualization) a graded-skew-symmetric multilinear operation $l_n\colon \wedge^n V\ra V $ of $\ZZ$-degree $2-n$. The classical master equation  for $S$ corresponds to a sequence of quadratic relations satisfied by the operations $l_n$:
\begin{multline*}
\hspace{-0.5cm} \sum_{r,s\geq 0\; \mr{s.t.} \; r+s=m}\; \sum_{\sigma\in \mr{Sh}_{r,s} } \pm l_{s+1}(l_r(x_{\sigma(1)},\ldots, x_{\sigma(r)}),\\
 ,x_{\sigma(r+1)},\ldots,x_{\sigma(m)}) =0 
\end{multline*}
---the ``homotopy Jacobi identities;'' here $m\geq 1$ enumerates relations and $x_1,\ldots,x_m\in V$ are any elements; the inner sum is over $(r,s)$-shuffles. Thus, the operations $l_n$ equip $V$ with the structure of an $L_\infty$ algebra (in particular, $l_1\colon V\ra V$ squares to zero and is a differential on $V$).\footnote{See, e.g., \cite{LS}. An alternative (equivalent) definition of an $L_\infty$ algebra is: a degree $1$ coderivation $Q$ of the augmented symmetric coalgebra $S^{\geq 1} (V[1])$ satisfying $Q^2=0$.} The inner product $\langle,\rangle$ is invariant for this $L_\infty$ structure, in the sense that the ``cyclic $L_\infty$ operations'' $\langle -, l_n(\cdots)\rangle\colon V\otimes \wedge^n V\ra \RR$ factor through $\wedge^{n+1}V$. One calls the data $(V,\langle,\rangle, \{l_n\}_{n\geq 1})$ a \emph{cyclic $L_\infty$ algebra} (of degree $-3$). 

The solution (\ref{S=S2+S3+..}) of the CME corresponding to a given cyclic $L_\infty$ algebra is
$$ S(\alpha)=\sum_{n\geq 1} \frac{1}{(n+1)!} \langle s(\alpha),l_n(s(\alpha),\ldots,s(\alpha))  \rangle  $$
with $\alpha\in \X$ and $s\colon \X\ra V$ the degree-shift (or ``suspension'') map.\footnote{
The picture of solutions of the master equation as corresponding to $L_\infty$ algebras and the idea of effective BV actions as corresponding to homotopy transfer is, to our knowledge, due to Andrey Losev \cite{Losev06, Losev_Berezin, Losev_IPMU} (precursors of this picture appeared in \cite{Zwiebach,Stasheff}). This idea was fleshed out in various examples in \cite{Krotov_Losev,AKLL} ($D=10$, $\mc{N}=1$ super Yang-Mills),  \cite{Mnev06, Mnev08, CMRcell} ($BF$ theory transferred to cochains of a triangulation), \cite{CM08,1dCS}  (Chern-Simons theory), \cite{CMRpert} (general gauge theories). 
}

\begin{example}\label{ex: CS from de Rham}
    For $M$ a closed 3-manifold and $\mathfrak{g}$ a quadratic Lie algebra (with invariant inner product $\langle,\rangle_\g$), one has a dg Lie algebra $\Omega^\bt(M)\otimes \g$ with inner product $\int_M \langle -\stackrel{\wedge}{,} -\rangle_\g$. In particular, it is a cyclic $L_\infty$ algebra of degree $-3$ with only operations $l_1=\ddd$ and $l_2=[\ ,\ ]$ nontrivial. The corresponding solution of the CME is the BV action of Chern--Simons theory
    \begin{equation}\label{e:CSAKSZ}
    S=\int_M \left(\frac12 \langle  A\stackrel{\wedge}{,} \ddd A \rangle_\g +\frac16 \langle  A\stackrel{\wedge}{,} [A,A] \rangle_\g\right)
    \end{equation} 
    where $A\in \Omega^\bt(M)\otimes \g[1]$ is the BV field---a nonhomogeneous $\g$-valued form on $M$.
\end{example}

\begin{remark}
    Given a solution of the quantum master equation $S_\hbar =S+\hbar S^{(1)}+\cdots$ on $\X$, one can reinterpret its monomials as a 2-parametric family of multilinear operations $c_n^{(l)}\colon \wedge^n V\ra \RR$, where $n$ is the ``arity'' (polynomial degree of the monomial) and $l$ is the ``loop number'' (power of $\hbar$ by which the monomial is accompanied). Due to QME for $S_\hbar$, this collection of operations satisfies a two-parametric family of quadratic relations. The structure $(V,\langle,\rangle,\{c_n^{(l)}\})$ is the \emph{quantum} (or ``loop-enhanced'') cyclic $L_\infty$ algebra. 
\end{remark}

\textbf{Fiber BV integrals and homotopy transfer.}
Assume that $S_\hbar=S+\hbar S^{(1)}+\cdots $ is a solution of the quantum master equation with $S$ of the form (\ref{S=S2+S3+..}). Assume that $\X$ is split into symplectically orthogonal subspaces $\X=\X_1\oplus \X_2$, and assume that the corresponding decomposition of the shifted space $V=V_1\oplus V_2$ is a splitting of a cochain complex (w.r.t.\ $l_1$) into subcomplexes, with $V_2$ acyclic. Choose a chain homotopy $K\colon V\ra V$ between $\mr{id}_V$ and $\mr{proj}_{V_1}$. One can then evaluate the fiber BV integral (\ref{fiber BV integral}) with gauge-fixing Lagrangian $\mc{L}=\mr{im}(K)\subset \X_2$. It yields an effective BV action $S_{1,\hbar}=S_1+S_1^{(1)}\hbar+\cdots$ which is, by Theorem \ref{thm: fib BV int}, a solution of the QME on $\X_1$. Perturbative evaluation of the integral (\ref{fiber BV integral}) yields a formula for $S_{1,\hbar}$ as a sum over connected Feynman graphs with propagator being the chain homotopy $K$, internal vertices decorated by coefficients of $S_\hbar$ and leaves decorated by inputs in $\X_1$. In particular, $S_1$ (the $O(\hbar^0)$ part of the effective action) is given by a sum over trees.

In the language of algebras up-to-homotopy, $S$ defines a cyclic $L_\infty$ algebra $\mc{A}=(V,\langle\ ,\ \rangle,\{l_n\})$ 
on $V$ and $S_1$ corresponds to the cyclic $L_\infty$ algebra $\mc{A}_1=(V_1,\langle\ ,\ \rangle_1,\{l_n^1\})$ on $V_1$ induced from $\mc{A}$ via \emph{homotopy transfer}. Homotopy transfer expresses the induced operations $l_n^1$ as sums over Kontsevich-Soibelman trees (see \cite{KS}); these trees exactly correspond to the tree Feynman diagrams for $S_1$.

The averaging map 
$$
\begin{array}{ccc}
C^\infty(\X) &\ra & C^\infty(\X_1) \\
\mc{O} &\mapsto& 
\frac{\int_\LL e^{\frac{i}{\hbar}S_\hbar}\mc{O}\mu }{e^{\frac{i}{\hbar}S_{1,\hbar}}\mu_1}
\end{array}
$$
truncated to the order $O(\hbar^0)$ can be interpreted as the pullback by 
an $L_\infty$ quasi-isomorphism $\mc{A}_1\ra \mc{A}$.

\begin{example}
    In the setting of Example \ref{ex: CS from de Rham}, take $V_1=H^\bt(M)\otimes \g$---the de~Rham cohomology with coefficients in $\g$ (which can be included in $V$, e.g., as harmonic forms). The corresponding induced cyclic $L_\infty$ structure on $H^\bt(M)\otimes \g$ is as follows: $\langle\ ,\ \rangle_1$ is the Poincar\'e pairing, $l^1_1=0$, $l_2^1$ is the cup product tensored with Lie bracket in $\g$; operations $l_{\geq 3}^1$ are the Massey brackets on $\g$-valued cohomology.
\end{example}

\section{Field theory}\label{sec:FieldTheory}
So far we have been working with finite-dimen\-sional $SP$-manifolds, but the applications to classical and quantum field theory we have in mind require an appropriately defined infinite-dimen\-sional analog. Here we will attempt at giving a---necessarily incomplete---survey of approaches to generalize the finite-dimen\-sional scenario.

\subsection{Locality, and infinite-dimen\-sional geometry}
A first observation is that field theory is in its essence \emph{local}. This is the statement that, on the one hand, our spaces of fields are sections of fiber bundles over smooth manifolds (possibly noncompact and possibly with boundary, see below). 

On the other hand, locality requires that the data specifying the field theory should only depend on a finite number of derivatives of the field configurations (i.e.\ only on finite-order jets of sections). An elegant way to phrase locality goes through the variational bicomplex and jet evaluations \cite{Zuckermann,Anderson,DeligneFreed}.
Alternatively this can be phrased as the requirement that local data be specified by integrals of (products of) differential operators on the space of sections: 
\[
S=\int_M \mathcal{L}(\phi,\partial \phi, \partial\partial \phi,\dots))\mr{Vol}_M
\]
The two approaches are equivalent.\footnote{Observe however that integration requires care when $M$ is noncompact.}

A third aspect of locality in field theory is that the Lie group actions that encode the symmetries of the theory are local as well. This means that the Lie groups involved are \emph{typically} mapping groups $\mathcal{G}= C^\infty(M,G)$ for some (finite-dimen\-sional) Lie group $G$---with the remarkable outlier of $\mathcal{G}=\mr{Diff}(M)$ for gravitational theories.

This means that we can consider both field configurations and their symmetry generators on equal footing, and construct the associated BV data in terms of sections of a (graded) fiber bundle $F\to M$, i.e.\  we work in $\F=\Gamma(M,F)$.

When the space of fields $\F$ is as above, one can endow it with a smooth structure and a differential calculus that has many of the good properties of the finite-dimen\-sional scenario (but not all!). Indeed, the space of sections of a vector bundle over a compact manifold is a (nuclear) Fr\'echet space, and this approach can be generalized (with appropriate caveats) to $M$ noncompact and to vector bundles with (certain) infinite-dimen\-sional fibers \cite{KrieglMichor}.  

\subsection{Infinite-dimen\-sional BV formalism}
Let us look at the case of a (graded) vector bundle over a closed manifold $M$ for concreteness, and let $\F$ be its space of smooth sections. 

On $\F$ there is a \emph{weak}, ($-1$)-symplectic form $\omega$, which can also be thought of as a local, degree $-1$, bilinear pairing $\omega\colon T\F\times T\F \to \mathbb{R}$ that is injective in both entries. Equivalently, we can think of $\omega$ as a local $2$-form on $\F$ of degree $-1$ and whose associated map $\omega^\flat\colon T\F \to T^*\F$ is injective. (Here dualization can be understood in the sense of continuous linear functionals with the strong topology.\footnote{Note that $T^*\mathcal{F}$ cannot be a locally convex smooth manifold unless it is finite-dimen\-sional, but it can be given a convenient smooth structure \cite{KrieglMichor}.})
The classical BV operator $Q$ can be thought of as a local degree $1$ cohomological vector field on $\F$, or alternatively as a derivation on the algebra of local Hamiltonian functions, i.e. functions $F\in C_{\mr{loc}}^\infty(\F)$ such that there exists $X_F\in \mathfrak{X}(F)$ with $\iota_{X_F}\omega = \delta F$. In particular $[Q,Q]=0$, and $Q$ itself is such that $\iota_Q\omega = \delta S$.

Gauge fixing in this generalized sense can be thought of as a choice of Lagrangian submanifold of the smooth infinite-dimen\-sional $(-1)$-symplectic (Fr\'echet) manifold $\F$. By this we mean an isotropic subspace $\LL$ with isotropic complement: $\omega\vert_{\LL} = \omega\vert_{\LL'} = 0$ and $\F = \LL \oplus \LL'$. (Observe that  in Sections \ref{sec:EffReg} and \ref{sec:BVBFV} one looks for such a Lagrangian submanifold only within a subspace of the space of fields, much like in the BV fiber integral picture of Section \ref{sec:BVfiberintegral}.)

\subsection{The AKSZ construction}
This is a general method that generates a space of fields endowed with a BV structure (odd symplectic form of ghost number $-1$ and even function of ghost number zero satisfying the CME) \cite{AKSZ}. The space of fields is defined as the mapping space $\Map(X,Y)$ between two graded manifolds $X$ and $Y$ with additional structures as explained below. 

The mapping space is a (usually infinite-di\-mensional Fr\'echet) graded manifold \cite{Roytenberg06} characterized by the property that for every finite-dimen\-sional graded manifold $Z$ the set of morphisms\footnote{A morphism between graded manifold is defined as a morphism between the sheaves that describe them.} from $Z$ to $\Map(X,Y)$ is naturally identified with the set of morphisms from $Z\times X$ to $Y$.\footnote{One can actually define the mapping space $\Map(X,Y)$ as the functor from the opposite category of finite-dimen\-sional graded manifolds $\mathbf{GrMfld}^\text{op}$ to the category of sets $\mathbf{Set}$ that sends $Z$ to $\text{Mor}(Z\times X,Y)$.
This way, $\Map(X,Y)$---which in category theory is called the internal hom from $X$ to $Y$---may be viewed as a generalized graded manifold, where by the latter we mean a functor $\mathbf{GrMfld}^\text{op}\to\mathbf{Set}$. An ordinary finite-dimen\-sional manifold $X$ is viewed in this context as the functor that sends $Z$ to $\text{Mor}(Z,X)$.} The body of $\Map(X,Y)$ is the infinite-dimen\-sional manifold of morphisms from $X$ to $Y$. 

Suppose we are given cohomological vector fields $Q_X$ and $Q_Y$ on $X$ and $Y$, respectively. They are naturally lifted to anticommuting cohomological vector fields $\Hat Q_X$ and $\Hat Q_Y$ on $\Map(X,Y)$ (roughly speaking, one identifies $Q_X$ and $Q_Y$ with their flows, which can be composed with a ``map'' from $X$ to $Y$ from the left of from the right). This gives rise to a cohomological vector field $Q=\Hat Q_X+\Hat Q_Y$ on $\Map(X,Y)$.

Next one assumes that $X$, of odd dimension $n$, is endowed with a Berezinian $\mu$. A function, or more generally a differential form, $\phi$ on $Y$ may be pulled back to $X\times \Map(X,Y)$
via the evaluation map $\text{ev}$. Now $\mathrm{ev}^*\phi$ can be integrated over $X$ via $\mu$ to a differential form $\mu_*\mathrm{ev}^*\phi$ on $\Map(X,Y)$ of the same form degree as $\phi$ but of internal degree and parity shifted by $-n$.

If $Y$ is endowed with a symplectic form of degree and parity $n-1$, then $\Hat\omega=\mu_*\mathrm{ev}^*\omega$ is an odd symplectic form of ghost number $-1$ on $\Map(X,Y)$.
If $S_Y$ is a hamiltonian function for $Q_Y$, it then turns out that $\Hat S_Y=\mu_*\mathrm{ev}^*S_Y$  is a hamiltonian function for $\Hat Q_Y$. In particular, it satisfies the CME.

Finally, one assumes that $\Hat Q_X$ is also hamiltonian, with hamiltonian function $S_X$. It then follows that $S=S_X+\Hat S_Y$ satisfies the CME.

A very important particular case is when $X=T^*[1]M$, where $M$ is an oriented $n$-dimensional manifold. In this case, $C^\infty(X)=\Omega^\bullet(M)$, and
one has natural data: $Q_X$ is the de~Rham differential and $\mu$ is the natural Berezinian: $\int_X f\mu=\int_M \Tilde f$, where $\Tilde f$ is $f$ interpreted as a differential form. It turns out that, if $\omega=\ddd\alpha$, then $S_X=\iota_{\Hat Q_X}\mu_*\mathrm{ev}^*\alpha$. In this example, the triple $(\Map(T^*[1]M,Y),\Hat\omega,S)$ are the BV data for a topological sigma model on $M$.

The description of the last example becomes more transparent if $Y$ has global Darboux coordinates $(p_i,q^i,\theta^\nu)$ 
with
\[
\alpha=\sum_i p_i\ddd q^i + \frac12\sum_\nu\theta^\nu\ddd\theta^\nu.
\]
In this case, $\Map(T^*[1]M, Y)$ has coordinates given by inhomogeneous
differential forms $P_i$, $Q^i$, and $\Theta^\nu$, with total degree and total parity (i.e., adding form degree and form parity to
ghost number and field parity) equal to the degree and parity of the corresponding target coordinate. The symplectic form and the action then read
\begin{align*}
    \Hat\omega &= \int_M\left(\sum_i \delta P_i\delta Q^i + \frac12\sum_\nu\delta \Theta^\nu\delta\Theta^\nu\right),\\
    S &= \int_M\Big( \sum_i P_i\ddd Q^i + \frac12\sum_\nu\Theta^\nu\ddd\Theta^\nu
\\ &\phantom{=\int_M} +S_Y(P,Q,\Theta)\Big).
\end{align*}
Particularly important outcomes of this construction are Chern--Simons theory,\footnote{See equation \eqref{e:CSAKSZ}.} $BF$ theory, and the Poisson sigma model.

\subsection{Effective regularization}\label{sec:EffReg}

A crucial ingredient for the (quantum) BV formalism is the BV Laplacian. When talking about infinite-dimen\-sional manifolds we see immediately that we cannot approach the problem by direct generalization of the finite-dimen\-sional case, owing to the fact that there is, generally, no appropriate (Bere\-zinian) measure on $\F$.

One way out is to define a family of (effective) theories for a set of parameters (usually energy or length), and define a BV Laplacian at each value of the parameter. This approach is based on\footnote{We also acknowledge Andrey Losev's contribution to the development of the perturbative, effective, quantization of field theory in the BV formalism, e.g.\ at the III G.A.P.\ meeting in Perugia \href{https://www.dmi.unipg.it/GAPIII}{https://www.dmi.unipg.it/GAPIII}.} \cite{Costello07,Costello}.

Let $S = \omega(\phi,Q\phi) + I(\phi)$ with $I(\phi)$ at least cubic, and assume that gauge fixing is given by $\LL=\mr{Im}(Q_{GF})$ for some odd operator $Q_{GF}\colon \F\to \F$ such that $Q_{GF}^2=0$ and $D=[Q,Q_{GF}]$ is an elliptic differential operator on $\F$. 
Assume furthermore that\footnote{Here we are using a slightly weaker notion of gauge fixing, with respect to the whole of $\F$. Observe however that $\mr{Im}(Q_{GF})$ is a true gauge fixing for the subspace $\F_2=\mr{Im}(Q_{GF})\oplus \mr{Im}(Q)$. This is an application of the BV fiber integral.}
\[
\F = \mr{Im}(Q)\oplus \mr{Im}(Q_{GF}) \oplus \mr{Ker}(D).
\]
One defines the integral kernel
\[
K_t * \phi := (-1)^{|K_t|}(1\otimes\omega)(K_t\otimes\phi) = e^{-tD}\phi,
\]
and we can write the propagator between two parameter scales $L_1$ and $L_2$ as 
\[
P_{L_1,L_2} = \int_{L_1}^{L_2}(Q_{GF}\otimes1)K_t\,dt,
\]
so that
\begin{align*}
P_{L_1,L_2} * \phi 
    &= \int_{L_1}^{L_2} Q_{GF}\,e^{-tD}\phi\,dt\\
    &= \int_{L_1}^{L_2} e^{-tD}Q_{GF}\phi\,dt.
\end{align*}
We can then define a regularized BV Laplacian at parameter $L$ by
\begin{multline}
\Delta_L = \frac{1}{2}\biggl(\int_{M\times M} K_L(x,y)\frac{\delta}{\delta\phi^i(x)}\frac{\delta}{\delta\phi^\dagger_i(y)} \\
+ \int_{M\times M} K_L(x,y)\frac{\delta}{\delta\phi^\dagger_i(x)}\frac{\delta}{\delta\phi^i(y)}\biggr).
\end{multline}

If we let the interaction term be ``effective'', i.e. we consider a family $\{I[L]\}$, we get the scale-$L$ quantum master equation
\[
(Q -i\hbar\Delta_L)e^{\frac{i}{\hbar}I[L]} = 0
\]

\subsection{Time ordering and BV Laplacians}
In certain cases one can construct a BV Laplacian following a slightly different approach. One starts from the data of classical observables, thought of as an (appropriately defined) commutative algebra of functions with a differential---the classical BV operator $Q$. Instead of deforming the differential to a quantum operator $Q\leadsto Q -i\hbar \Delta$, for some appropriately regularized BV Laplacian $\Delta$, one can instead deform the associative product into what is often called a \emph{time-ordered product}.\footnote{The name comes from the original applications to Lorentzian QFT, where the causal structure of space--time (hence the time ordering) has a crucial role.} Suppose that $G$ is a Green's function for the quadratic part of the (gauge fixed) action functional (a differential operator $P$ with additional regularity properties). Then we can define a time-ordering map as 
\[
\tau F = \exp\left(\frac{\hbar}{2}\left\langle G, \frac{\delta^2}{\delta \phi^2}\right\rangle \right)F,
\]
where $F$ belongs to some restricted class of ``regular'' local functionals. We define the time-ordered product as:
\[
F\cdot_\tau G = \tau(\tau^{-1} F \cdot \tau^{-1} G)
\]
One can extend the domain of definition of time ordered products after choosing a consistent choice of renormalization scheme, and it is possible to show that
\[
\hat{Q} \doteq \tau^{-1} \cdot Q \cdot \tau = Q - i\hbar \Delta_{\mathrm{ren}}
\]
where $\Delta_{\mathrm{ren}}$ is now a renormalized BV Laplacian. In concrete examples where $P$ is a normally hyperbolic operator, say the d'Alembertian on Minkowski space, $G$ is taken to be the Feynman propagator, and the time ordering is at first defined only on regular functionals. Via appropriate regularization, this definition can be extended to the larger class of microcausal functionals, and the resulting associative (quantum) algebra is given by formal power series with coefficients therein. A convenient choice of renormalization scheme is given by the Epstein--Glaser procedure \cite{EG}. See \cite{Rejzner} and \cite{RejznerEnc} for details.

\subsection{Field theory on manifolds with boundary}\label{sec:BVBFV} 
If space--time $M$ has a nonemtpy boundary, 
all the above has to be modified for the following (interrelated) reasons:
\begin{itemize}
    \item one needs boundary conditions to define the propagator;
    \item the functional integral is no longer supposed to give a normalization factor (the partition function) but a state (a ``function on the space of boundary conditions");
    \item the BV master equation is spoiled by boundary terms.
\end{itemize}

The BV-BFV formalism is a way to fix these issues \cite{CMR14,CMRpert}. Let us start with the last one. We focus for the moment on the classical master equation $(S,S)=0$, as the quantum version requires anyway a regularization to be discussed later. We assume the BV action $S$ and the BV form $\omega$ to be local functionals on the space of fields $\mathcal{F}$ and the classical master equation to be satisfied when there is no boundary. The classical BRST operator $Q=(S,\ )$---equivalently, $\iota_Q\omega=\delta S$---is then also local and satisfies $[Q,Q]=0$.

If there is a boundary, the equation $\iota_Q\omega=\delta S$ may be spoiled by boundary terms. We call $\Tilde{\alpha}$ the error term: $\Tilde\alpha\coloneqq\iota_Q\omega-\delta S$. Then we take its differential
$\Tilde\omega\coloneqq\delta\Tilde\alpha=-\LL_Q\omega$. The even two-form $\Tilde\omega$, of ghost number zero, plays the role of presymplectic structure on the space of boundary fields. In good cases, it has a regular kernel, and the quotient $\mathcal{F}^\partial$ by it is smooth (see \cite{CattaneoEnc} for further details on this procedure). It is equipped with a symplectic form $\omega^\partial$ such that $\pi^*\omega^\partial=\Tilde\omega$, where $\pi\colon\mathcal{F}\to\mathcal{F}^\partial$ is the canonical projection. It turns out that $Q$ is projectable to a uniquely determined vector fields $Q^\partial$ which satisfies $[Q^\partial,Q^\partial]=0$ and $\LL_{Q^\partial}\omega^\partial=0$. One should think of the triple $(\mathcal{F^\partial},\omega^\partial,Q^\partial)$ as the phase space of the theory (more precisely, functions on $\mathcal{F^\partial}$ with the differential $\LL_{Q^\partial}$ should be though of as a resolution of the functions on the reduced phase space).
The space $\mathcal{F}^\partial$ typically consists of fields, and some of their transversal jets, on the boundary. 

The (geometric) quantization of the theory should then proceed by first selecting a polarization (a foliation with Lagrangian leaves) of $\mathcal{F}^\partial$. For simplicity, we assume $\mathcal{F}^\partial=T^*\mathcal{B}$, for some choice of the base manifold $\mathcal{B}$, with vertical polarization.

The functional-integral quantization now requires a regularization. To start with, one has to regularize the space of fields $\mathcal{F}$ in a way relative to $\mathcal{B}$. Typically, one redefines $\mathcal{F}$ as a product $\mathcal{F}'\times\mathcal{B}$, where $\mathcal{F}'$ consists of fields with appropriate boundary conditions, relative to the choice of $\mathcal{B}$.\footnote{Usually, the product $\mathcal{F}'\times\mathcal{B}$ is not the original space $\mathcal{F}$ but only an appropriate regularization thereof.} Under good conditions, the space $\mathcal{F}'$ is a BV space, and one can apply the (field-theory version of the) BV formalism to compute $\psi\coloneqq\int_{\mathcal{L}} e^{\frac i\hbar S}$.\footnote{The choice of boundary conditions inherent to $\mathcal{F}'$ allows one in particular to define the propagator.}
As $S$ depends parametrically on $\mathcal{B}$, one can view $\psi$ as a state for the theory, with space of states an appropriately defined space of functions on $\mathcal{B}$. In several cases, one can show that this quantization satisfies the following desiderata:
\begin{itemize}
    \item The differential $\LL_{Q^\partial}$ gets quantized to an operator $\Omega$ that satisfies $\Omega^2=0$, so that one can define the quantization of the reduced phase space 
    as the cohomology of $\Omega$.
    \item The state $\psi$ is $\Omega$-closed, and a deformation of the gauge fixing changes it by an $\Omega$-exact term, so the class of $\psi$ is a state for the reduced space.
    \item If we cut the space--time manifold $M$ a\-long a hypersurface $\Sigma$ into components $M_1$ and $M_2$, we can get the partition function for $M$ (or the state, if $M$ has boundary) as the pairing of the states for the two components $M_1$ and $M_2$.
\end{itemize}

\section{Conclusions}
In this note we have overviewed the BV formalism, its geometric interpretation, and its various implementations in field theory.


\begin{thebibliography}{99}
\bibitem{1dCS} A. Alekseev, P. Mnev. ``One-dimensional Chern-Simons theory.'' Commun. Math. Phys. 307.1 (2011) 185--227.
\bibitem{AKSZ} M. Alexandrov, M. Kontsevich, A. Schwarz, O. Zaboronsky, ``The geometry of the master equation and topological quantum field theory.'' International Journal of Modern Physics A 12.07 (1997): 1405--1429.

\bibitem{AKLL} V. Alexandrov, D. Krotov, A. Losev, V. Lysov. ``On pure spinor superfield formalism.'' JHEP 2007.10 (2007) 074. 
\bibitem{Anderson} I. Anderson, Introduction to the variational bicomplex, in Mathematical aspects of classical field theory, Contemp. Math. 132 (1992) 51–73.
\bibitem{Anselmi} D. Anselmi, ``Removal of divergences with the Batalin--Vilkovisky formalism.'' Class. Quantum Grav. 11 (1994): 2181--2204.

\bibitem{BBH95} G. Barnich, F. Brandt, M. Henneaux. ``Local BRST cohomology
in the antifield formalism. I. General theorems.'' Commun. Math. Phys.
174 (1995), 57--91.



\bibitem{BV81} I. A. Batalin,  G. A. Vilkovisky, ``Gauge algebra and quantization.'' Physics Letters B 102.1 (1981): 27--31.
\bibitem{BV83} I. A. Batalin,  G. A. Vilkovisky,
``Quantization of gauge theories with linearly dependent generators.'' Physical Review D 28.10 (1983): 2567.
\bibitem{Bat79} M. Batchelor, “The Structure of Supermanifolds”, Transactions of the
American Mathematical Society 253, pp. 329–338 (1979).
\bibitem{BRS} C. Becchi, A. Rouet, R. Stora, ``Renormalization of gauge theories.'' Ann. Phys. 98.2 (1976): 287--321.

\bibitem{Encyclopedia} M. Bojowald and R. J. Szabo (eds.), Encyclopedia of Mathematical Physics.

\bibitem{CanepaCattaneoSchiavinaAKSZ}
G. Canepa, A. S. Cattaneo, M. Schiavina, ``General relativity and the AKSZ construction.” Commun. Math. Phys. 385 (2021): 1571--1614.

\bibitem{SimaoCattaneoSchiavina}
F. M. Castela Sim\~ao, A. S. Cattaneo, M. Schiavina, ``BV equivalence with boundary,” arXiv:2109.05268; to appear in Lett. Math. Phys.



\bibitem{CattaneoEnc} A. S. Cattaneo, ``Phase space for gravity with boundaries'', Encyclopedia of Mathematical Physics.

\bibitem{CF} A. S. Cattaneo, G. Felder, ``A path integral approach to the Kontsevich quantization formula.'' Commun. Math. Phys. 212 (2000): 591--611.

\bibitem{CM08} A. S. Cattaneo, P. Mnev, ``Remarks on Chern–Simons invariants.'' Commun. Math. Phys. 293.3 (2010) 803--836.

\bibitem{CMR14} A. S. Cattaneo, P. Mnev, N. Reshetikhin, ``Classical BV Theories on Manifolds with Boundary.'' Commun. Math. Phys. 332, pages 535–603 (2014).

\bibitem{CMRpert} A. S. Cattaneo, P. Mnev, N. Reshetikhin, ``Perturbative quantum gauge theories on manifolds with boundary.'' Commun. Math. Phys. 357.2 (2018) 631--730.

\bibitem{CMRcell} A. S. Cattaneo, P. Mnev, N. Reshetikhin, ``A cellular topological field theory.'' Commun. Math. Phys. 374.2 (2020) 1229--1320. 

\bibitem{CattaneoRossi} A. S. Cattaneo, C. Rossi, ``Wilson surfaces and higher dimensional knot invariants.'' Commun. Math. Phys. 256 (2005): 513–-537.
\bibitem{Chas-Sullivan} M. Chas, D. Sullivan, ``String topology.'' arXiv preprint math/9911159 (1999).
\bibitem{ChiaffrinoSachs} C. Chiaffrino, I. Sachs, ``QFT with stubs''. J. High Energ. Phys. 2022, 120 (2022).

\bibitem{Costello07} K. J. Costello, ``Renormalization and the Batalin--Vilkovisky formalism,'' arXiv preprint 0706.1533.

\bibitem{Costello} K. J. Costello, ``Renormalization and Effective Field Theory,'' Mathematical surveys and monographs, Volume 170, American Mathematical Society, 2011.

\bibitem{EG} H. Epstein, V. Glaser, `` The role of locality in perturbation theory.'' Annales Henri Poincar\'e 19 (3) 211--295 (1973).

\bibitem{DeligneFreed} P. Deligne, P. Etingof, D. Freed, L. Jeffrey, D. Kazhdan, J. Morgan, D. Morrison and E. Witten, eds. ``Quantum Fields and Strings, A course for mathematicians,'' 2 vols. Amer. Math. Soc. Providence 1999. 

\bibitem{FP} 
L. D. Faddeev, V. Popov,  ``Feynman diagrams for the Yang-Mills field.'' Physics Letters B 25.1 (1967) 29.
\bibitem{FK} G. Felder, D. Kazhdan, T. M. Schlank, ``The classical master equation.'' Perspectives in representation theory, Contemp. Math 610 (2014): 79--137.
\bibitem{FHST} J. Fisch, M. Henneaux, J. Stasheff, C. Teitelboim, ``Existence, uniqueness and cohomology of the classical BRST charge with ghosts of ghosts''. Commun. Math. Phys. 120, 379–407 (1989).

\bibitem{Getzler}  E. Getzler, ``Batalin-Vilkovisky algebras and two-dimensional topological field theories.'' Communications in mathematical physics 159.2 (1994): 265--285.

\bibitem{Henneaux} M. Henneaux, ``Elimination of the auxiliary fields in the antifield formalism.'' Phys. Lett. B 238 (1990): 299--304.

\bibitem{HenneauxEnc} M. Henneaux, ``BRST Quantization'' in Encyclopedia of Mathematical Physics.

\bibitem{HenneauxTeitelboim} M. Henneaux, C. Teitelboim, ``Quantization of gauge systems'', Princeton University Press, (1992).

\bibitem{Khudaverdian91} O. M. Khudaverdian, ``Geometry of superspace with even and odd brackets,'' J. Math. Phys. 32, 1934–-1937, (1991).
\bibitem{Khudaverdian}  H. M. Khudaverdian, ``Semidensities on odd symplectic supermanifolds.'' Commun. Math. Phys.
247.2 (2004) 353--390.

\bibitem{KS} M. Kontsevich, Y. Soibelman. ``Homological mirror symmetry and torus fibrations.'' In \textit{Symplectic geometry and mirror symmetry.} 2001. 203-263.

\bibitem{KrieglMichor} A. Kriegl, P. W. Michor, ``The Convenient Setting of Global Analysis'', American Mathematical Soc., 1997.

\bibitem{Krotov_Losev} D. Krotov, A. Losev. ``Quantum field theory as effective BV theory from Chern–Simons.'' Nucl. Phys. B 806.3 (2009) 529--566.

\bibitem{YKSGersten} Kosmann-Schwarzbach, Y. ``Exact Gerstenhaber algebras and Lie bialgebroids.'' Acta Appl Math 41, 153–165 (1995).

\bibitem{LS} T. Lada, J. Stasheff. ``Introduction to SH Lie algebras for physicists.'' arXiv preprint hep-th/9209099 (1992).

\bibitem{Losev06} A. Losev, “BV formalism and quantum homotopical structures,” Lectures at GAP3, Perugia,
2006.

\bibitem{Losev_Berezin} A. Losev, "From Berezin integral to Batalin-Vilkovisky formalism: a mathematical physicist's point of view.''
 In \textit{Felix Berezin: life and death of the mastermind of supermathematics,} pp. 3--30. 2007.

\bibitem{Losev_IPMU} A. Losev, ``TQFT, homological algebra and elements of K. Saito's theory of Primitive form: an attempt of mathematical text written by mathematical physicist.'' In \textit{Primitive Forms and Related Subjects -- Kavli IPMU 2014}, pp. 269--293. Mathematical Society of Japan, 2019.

\bibitem{Mnev08} P. Mnev, ``Discrete BF theory,'' arXiv:0809.1160 [hep-th]. 

\bibitem{Mnev06} P. Mnev, ``Notes on simplicial $BF$ theory,'' 
Moscow Mathematical Journal 9 (2009): 
371--410.

\bibitem{Mnev15} P. Mnev, ``A construction of observables for AKSZ sigma models.''
Lett. Math. Phys. 105 (2015): 1735--1783.

\bibitem{Moshayedi} N. Moshayedi, ``Formal global AKSZ gauge observables and generalized Wilson surfaces.'' Annales Henri Poincar\'e 21 (2020): 2951--2995. 

\bibitem{Rejzner} K. Rejzner, ``Perturbative Algebraic Quantum Field Theory: an introduction for mathematicians", Springer International Publishing (2016).

\bibitem{RejznerEnc} K. Rejzner, Encyclopedia of Mathematical Physics.

\bibitem{RocekZeitlin} M. Rocek, A. M. Zeitlin, ``Homotopy algebras of differential (super)forms in three and four dimensions.'' Lett. Math. Phys. 108 (2018): 2669--2694. 

\bibitem{Roytenberg06} D. Roytenberg, ``AKSZ-BV formalism and Courant algebroid-induced topological field theories.'' Lett. Math. Phys. 79 (2007): 143--159. 


\bibitem{Schwarz} A. Schwarz,  ``Geometry of Batalin-Vilkovisky quantization.'' Communications in Mathematical Physics 155.2 (1993): 249--260.
\bibitem{Severa} P. \v{S}evera, ``On the origin of the BV operator on odd symplectic supermanifolds.'' Lett. Math.
Phys. 78.1 (2006) 55--59.
\bibitem{Stasheff} J. Stasheff,  ``The (secret?) homological algebra of the Batalin-Vilkovisky approach.'' Contemporary mathematics 219 (1998): 195--210.
\bibitem{T} I. V. Tyutin, ``Gauge invariance in field theory and statistical physics in operator formalism.'' Lebedev Physics Institute preprint 39 (1975), arXiv:0812.0580.
\bibitem{Vinogradov} Alexandre Vinogradov, A spectral sequence associated with a non-linear differential equation, and the algebro-geometric foundations of Lagrangian field theory with constraints , Sov. Math. Dokl. 19 (1978) 144–148.
\bibitem{Witten} E. Witten, ``A note on the antibracket formalism.'' Modern Physics Letters A 5.07 (1990): 487--494.
\bibitem{ZinnJustin} J. Zinn-Justin, ``Renormalization of gauge theories.'' In: Rollnik, H., Dietz, K. (eds) Trends in Elementary Particle Theory. Lecture Notes in Physics, vol 37. Springer, Berlin, Heidelberg, 1975. Pages 1--39.
\bibitem{Zwiebach} B. Zwiebach, ``Closed string field theory: Quantum action and the Batalin-Vilkovisky master equation,'' Nucl. Phys. B 390 (1993), 33--152.

\bibitem{Zuckermann} G. 
Zuckerman, ``Action principles and global geometry,'' in: Shing-Tung Yau (ed.) Mathematical Aspects of String Theory, World Scientific (1987) 259-284.
\end{thebibliography}
\end{document}